\def\gapprox{{_>\atop{^\sim}}}
\def\lapprox{{_<\atop{^\sim}}}

\def\cmmd{\rm {cm^{-3}}}
\def\cmmt{\rm {cm^{-2}}}

\def\s-1{\rm {s^{-1}}}
\def\twco{$^{12}$CO}
\def\thco{$^{13}$CO}

\def\sskip{\vskip \baselineskip \noindent}

\def \0{\hbox to 0.5 em{}}
\def \etal{{\it et al.\ }}

\def \kms{km s$^{-1}\ $}

\def\myoversim#1#2{\lower2.0pt\vbox{\baselineskip0pt \lineskip0.2pt
    \ialign{${\mathsurround=0pt }#1\hfil##\hfil$\crcr
    #2\crcr\sim\crcr}}}

\documentstyle[12pt,aasms4]{article}

\received{1996 August 8}
\accepted{1996 November 13}

\slugcomment{To appear in 
The Astrophysical Journal Letters, 1997 February 1}

\begin{document}

\title{{.}\break \break
Variation of Molecular Line Ratios and\break
Cloud Properties in the Arp~299 Galaxy Merger}

\author{
S. Aalto
}\affil{
Division of Physics, Mathematics \& Astronomy, \\
Caltech 105-24, Pasadena CA 91125, USA\\
sab@caltech.edu}

\author{
Simon J. E. Radford
}\affil{
National Radio Astronomy Observatory, \\
949 North Cherry Avenue, Tucson, AZ 85721-0665\\
sradford@nrao.edu}

\and

\author{
N. Z. Scoville and  A. I. Sargent
}\affil{
Division of Physics, Mathematics \& Astronomy, \\
Caltech 105-24, Pasadena CA 91125, USA\\
nzs@caltech.edu, afs@caltech.edu}

\begin{abstract}

High resolution observations of \twco ($2.''3$),
\thco ($3.''9$), and HCN ($5.''4$) $J$=1--0 in the galaxy merger Arp~299
(IC~694 and NGC~3690) show the line ratios vary dramatically
across the system. The \twco/\thco\ ratio is unusually large,
$60 \pm 15$, at the IC~694 nucleus,
where \twco\ emission is very strong, and much smaller, $10 \pm 3$, in the
southern extended disk of that galaxy.
Elsewhere, the \twco/\thco\ line ratio is 5-20, typical of spiral galaxies.
The line ratio variation in the overlap
between the two galaxies is smaller, ranging from $10 \pm 3$ in the east
to $20 \pm 4$ in the west.

The \twco/HCN line ratio also varies across Arp~299, although to a lesser degree.
HCN emission is bright towards each galaxy nucleus and in the extranuclear
region of active star formation; it was not detected in the IC~694 disk, or
the eastern part of the overlap region, leading to lower limits of 25 and 20
respectively. By contrast, at the
nuclei of IC~694 and NGC~3690 the ratios are $9 \pm 1$ and $14 \pm 3$
respectively. In the western part of the overlap region it is $11 \pm 3$.

The large \twco/\thco\ 1--0 intensity ratio at the nucleus of IC~694 can
primarily be attributed to a low to moderate optical depth ($\tau \lapprox 1$) in the
\twco\ 1--0 line. These data support the hypothesis that unusually high
\twco/\thco\ line ratios ($>$ 20) are associated with extremely compact molecular
distributions in the nuclei of merging galaxies. Relative to \twco,  \thco\ 1--0 is brightest
in quiescent regions of low \twco\ surface brightness and
weakest in starburst regions and the galactic nuclei.
A medium consisting of dense ($n=10^4 - 10^5$ $\cmmd$)
and warm ($T_{\rm k} > 50$ K) gas will
reproduce the extreme line ratios observed in the nucleus of
IC~694, where the area filling factor must be at least 20\%.

\end{abstract}

\keywords{
    galaxies: evolution
--- galaxies: individual(Arp\,299)
--- galaxies: ISM
--- galaxies: starburst
--- radio lines: galaxies
--- radio lines: ISM }

\section{Introduction}

The inner kiloparsecs of starburst and interacting galaxies harbor
stunning amounts of molecular gas, $10^9 - 10^{10} M_{\odot}$ (e.g.,
Scoville \etal\ 1991; Bryant \& Scoville 1996). In these environments,
molecular clouds are subject to intense radiation fields,
supernovae explosions, winds from newborn hot stars, strong tidal forces, and
gas surface densities several order of magnitudes higher than in
the Milky Way disk. These are also
extremely active star formation sites. Knowledge of the physical conditions
and structure of molecular gas in interacting systems
is essential to understand the starburst activity and its role in
galaxy evolution.

Arp~299 is an IR-luminous ($L_{\rm IR} \approx 8\times10^{11}$ L$_{\odot}$)
merger system of two galaxies, IC~694 and NGC~3690.
Strong \twco\ emission has been detected from the nuclei of IC~694 and NGC~3690
and from the interface between the two galaxies
(Solomon \& Sage 1988; Casoli \etal 1989; Sargent \etal\ 1987; Sargent
\& Scoville 1991).
The two nuclei, as well as the western overlap region,
currently harbor intense star formation activity (c.f., Gehrz \etal\ 1983;
Baan \& Haschick 1990). Furthermore, the nucleus of IC~694 is
a flat-spectrum radio source, and may be an AGN (Gehrz \etal\ 1983).

Lower resolution (single dish) observations reveal an unusually large \twco/\thco\ 1--0 line
intensity ratio, $\gapprox 20$ in Arp~299 (Aalto \etal\ 1991; Casoli \etal\
1992). These observations left it unclear whether this is due to weak \thco\
in the whole system or to a varying \twco/\thco\ 1--0 line ratio.
Observations at 20$''$ and 11$''$ resolution by Casoli \etal (1992) suggest little variation
in  the \twco/\thco\ 1--0 line ratio and none in the \twco/\thco\ 2--1 line ratio.
In contrast, Aalto \etal (1995) note substantial variations at 28$''$ resolution
in the \twco/\thco\ 2--1 ratio: the
ratio is about 17 in IC~694; close to 9 in NGC~3690; and about 7 in the interface
region between the two disks.
Solomon \etal\ (1992) detected bright HCN emission in 28$''$ maps of Arp~299.
They measured \twco/HCN ratios of 11 in IC~694 and 13 in the interface
region.

\section {Observations and Results}

Aperture synthesis \twco, \thco, and HCN 1--0 mapping of Arp~299 was carried out
with the Owens Valley Radio Observatory (OVRO) millimeter array between March, 1995, and
February, 1996. SIS receivers on the six 10.4 m telecopes provided typical
system temperatures (SSB) of 600 K, 450 K, and 350 K for \twco, \thco, and HCN.
Quasars 1150+497 and 0917+449 were used for phase calibration
and Uranus and Neptune for absolute flux calibraton.
The synthesised beams are $2.''5 \times 2.''2 $ for \twco\ (uniform weighting),
$4.''3 \times 3.''6 $ for  \thco\ (natural weighting), and $5.''6 \times
5.''3$ for HCN (natural weighting).
At 2.6 mm wavelength with $2.''3$ resolution, a brightness
temperature of 1 K corresponds to 57 mJy beam$^{-1}$.
The digital correlator, centered at $V_{\rm LSR} = 3100$ \kms, provided
a total velocity coverage
of  1123 \kms\ for \twco, 1175 \kms\ for \thco, and 1407 \kms\ for HCN.
Data were binned to 4 MHz resolution, corresponding to
10 \kms\ for \twco\ and \thco\ and 13 \kms\ for HCN.
At 110 GHz an unresolved continuum
source of $17 \pm 2$ mJy was detected in the center of IC~694. Continuum
emission was also detected in NGC~3690 ($5 \pm 2$ mJy) and the overlap region
($9 \pm 2$ mJy). We subtracted this continuum emission from
the line emission before maps were made.

The main structures found by Sargent \& Scoville (1991)
with the three telescopes array are recovered in our new \twco\ map (Figure 1a),
but our increased $uv$ coverage enable improved deconvolution.
Lower surface brightness emission (A2), possibly a molecular disk or bar,
extends 10-15$''$ southeast of the IC~694 nucleus (A1),
coincident with the remnant optical disk. The center of NGC~3690 (B1) is also
surrounded by weak, extended emission (B2 and B3) with a somewhat S-shaped morphology.
Where the two galaxies overlap, three distinctive clumps (C1, C2, and C3)
can now be discerned. Weaker, extended emission is also recovered better in our new \twco\
map and clumpy structures can be distinguished at the center of the map (F).
These structures appear to connect the major components A, B, and C.
In addition, there is a clump (D) north of the main
features with systemic velocity $\approx$ 3280 \kms, which is beyond
the bandwidth of the earlier OVRO data.

Bright 6 cm radio continuum peaks at the nuclei of IC~694 and NGC~3690
(Gehrz \etal\ 1983; Condon \etal\ 1991) coincide with the \twco\ peaks to within the estimated positional
uncertainty ($0.''5$). There is also reasonable positional agreement between the two
brightest \twco\ clumps in the overlap region (C) and two additional 6 cm radio continuum
peaks: C1 and the western radio continuum peak are also coincident within $0.''5$, although
the discrepancy between C2 and the eastern radio continuum peak is
somewhat larger, $\Delta \alpha = 0.''8$.
Weaker radio continuum emission at 18 cm and 6 cm is spatially coincident with the
extended \twco\ emission in regions A2, B2, B3, and F (Baan \& Haschick 1990; Gehrz \etal\ 1983).

The velocity field image (Figure 2, Plate 1) reveals a monotonic shift from
the blueshifted emission from A2 in IC~694 to redshifted emission from the
overlap region C and from region D. Velocity gradients within C and D are small.
The velocity field of NGC~3690 is complicated by a double-peaked emission
structure of B2, and both B2 and B3 appear blueshifted relative to the center, B1.

The nuclei of both IC~694 (A1) and NGC~3690 (B1) remain unresolved by our $2.''2$
synthesised beam. A two-dimensional Gaussian fit to the nuclear emission of IC~694
and NGC~3690 yields upper limits to the source diameters of $1.''4$ and $1.''5$, corresponding
to radii of 140 and 150 pc, respectively (for $D$=42 Mpc). This implies a lower limit of 18 K to
the \twco\ brightness temperature in the IC~694 nucleus, and therefore,
the cloud filling factor is quite high in the inner
200 pc. Even if the intrinsic brightness temperature is as high as 100 K,
the surface filling factor of clouds is still almost 20\%. The three features in the overlap
region are resolved, with sizes of $792 \times 322$ pc (C1),
$876 \times 428$ pc (C2), and $684 \times 456$ pc (C3). Derived properties for all
designated regions are presented in Table 1.

The total molecular mass of Arp~299, estimated from
$M({\rm H}_2)=1.2 \times 10^4 S \Delta v D^2$ M$_{\odot}$ ($D$ is the distance in Mpc; $S \Delta v$ is the
integrated \twco\ 1--0 line flux in Jy K kms$^{-1}$), is $7.5 \times 10^9$ M$_{\odot}$, 87\% of
the value estimated by Solomon and Sage (1988) from their single dish map. This formula
corresponds to $N$(H$_2$)/$I$(\twco)=$3 \times 10^{20}$ $\cmmt$ (K \kms )$^{-1}$
(Sargent \& Scoville 1991), a standard Galactic \twco\ luminosity to H$_2$ mass ratio.
The conversion factor may vary, however, across Arp~299, since the line ratios indicate
different cloud properties.

The most striking feature of the \thco\ 1--0 map (Figure 1b) is the {\it absence of
strong emission at the nucleus of IC~694 (A1)}. Emission {\it is} detected in the
A2 region of IC~694, in the overlap region C, and in NGC~3690.
Unlike Casoli \etal\ (1992) we find significant variation in the \twco/\thco\
1--0 line ratio across Arp~299.
>From the very high value of $60 \pm 15$ at the IC~694 nucleus (A1) the ratio drops
to $10 \pm 3$ in the A2 region.
The global ratio for NGC~3690 is $13 \pm 2$, but there is an indication that the ratio
is higher in the nucleus, B1, than in B2 or B3. The ratio for the east (C3) and
west (C1) overlap region is $11 \pm 3$ and $20 \pm 4$ respectively. The spatial
correlation between the \thco\ emission and radio continuum emission peaks is also poor.

As for \twco, the HCN map (Figure 1c) is dominated by a peak at the center of
IC~694 (A1), where the \twco/HCN line ratio is $9 \pm 1$. Emission is also detected
at the nucleus
of NGC~3690, where the ratio is $14 \pm 3$, and in the western (C1)
overlap region, where \twco/HCN= $11 \pm 3$. There is very little HCN in regions
A2 and C3. The HCN and \thco\ line emission peaks appear anticorrelated.

\section {The molecular line-ratios}

As a system, Arp~299 is not deficient in \thco.
In all regions but the core of IC~694, we observe \twco/\thco\ line ratios
typical of star forming regions
in other galaxies (e.g., Aalto \etal\ 1995, 1991; Young \& Sanders 1986;
Rickard \& Blitz 1985). Relative to \twco, \thco\ 1--0
is brightest in quiescent regions of low \twco\ surface brightness, and
weak in starburst regions and galactic nuclei. In contrast, HCN 1--0, like \twco,
is bright in the two galaxy centers and in regions of active star formation.
These line ratio variations are, most likely, caused by differences in
line excitation. Our results support the suggestion (Aalto \etal\ 1995) that
{\it unusually} high (i.e. $>20$) \twco/\thco\ 1--0 line ratios tend to be
associated with extremely compact molecular distributions centered on the nuclei
of merging galaxies and are primarily due to a small or moderate optical depth,
$\tau \lapprox 1$, in the \twco\ 1--0 line. High ambient
pressures, strong tidal forces and ongoing starburst or AGN activity lead to substantial
changes in cloud structure and physical conditions.
Bryant (1996) also finds that HCN 1--0 is bright in regions of high \twco\ surface brightness
in merging galaxies.

\subsection{IC~694}

The line ratio variation from the IC~694 disk, where \twco/\thco\ = 10 and
\twco/HCN $\gapprox$ 25, to its nucleus, where the corresponding values are
60 and 9, reflects a dramatic change in cloud properties.

The bright HCN line accompanied by relatively weak \thco\ emission (HCN/\thco\ = $7 \pm 2$)
implies a population of unusually dense and warm clouds. The
HCN 1--0 strength implies densities $n \gapprox 10^4$ $\cmmd$ if
the HCN excitation is dominated by collisions with H$_2$. It is also likely that the density is
$\lapprox 10^5$, so that most of the HCN population will remain in the lower
levels and $\tau_{10} > 1$. At these densities, the
\twco\ and \thco\ 1--0 transitions are thermalised. If the kinetic temperature is also
high, the lower levels may become significantly depopulated,
effectively reducing the optical depth of the 1--0 line.
Then the \thco\ 2--1/1--0 line ratio should be $>$ 1.
Comparing the single dish \thco\ 2--1 flux (Aalto \etal\ 1995) with our \thco\ 1--0
flux from the nucleus of IC~694, we estimate \thco\ 2--1/1--0 $\gapprox 2$,
implying that the gas temperature is high,  $>$ 50 K.
Although the single dish beam was large, 28$''$, the bulk ($ \gapprox $65\%) of the
\twco\ 2--1 emission within the beam originates in the nucleus of IC~694 (A1).

Since the lower transitions of \twco\ and \thco\ appear
thermalised, LTE can be assumed to estimate the \twco\ column density, $N$(\twco).
We find it unlikely that the \twco/\thco\ abundance ratio is extremely high (section 3.3).
Therefore, a high \twco/\thco\ 1--0
line ratio indicates a low to moderate optical depth ($\tau \lapprox 1$) in the \twco\ 1--0 line.
The high intrinsic \twco\ 1--0 brightness temperature makes $\tau << 1$ unlikely, and we
therefore assume $\tau_{10}(^{12}{\rm CO}) \approx 1$. The optical depth of the 1--0 transition
can be expressed as $\tau_{10}(^{12}{\rm CO}) \approx 3.9 \times 10^{-15} N(^{12}{\rm CO})
(1- e^{-5.53/T_{\rm ex}})/T_{\rm ex} \Delta V$.
For a temperature $T_{\rm ex}$=100 K, line width $\Delta V = 5$ \kms, and $\tau_{10}$=1,
the \twco\ column density $N(^{12}{\rm CO}) = 2 \times 10^{18}$ $\cmmt$ (per cloud) and the resulting
brightness temperature $T_B(^{12}{\rm CO}) \approx 60$ K. For a density of $n= 10^4$ $\cmmd$
and a \twco\ abundance, [\twco/H$_2$]=$5 \times 10^{-5}$, the cloud
radius $r=N(^{12}{\rm CO})/2x(^{12}{\rm CO})n({\rm H}_2)$=0.7 pc --- not unlike cores or clumps within
Giant Molecular Clouds in the Galaxy. At this gas density, HCN is not thermalised so
$T_{\rm ex}({\rm HCN})$ for the 1--0 line will be significantly lower
than that for \twco. If the abundance ratio [\twco/HCN] $\approx 10^3$, then $\tau_{10}$(HCN) is
an order of magnitude higher than $\tau_{10}$(\twco). The resulting $T_{\rm B}$(HCN) is sufficiently
high to account for a \twco/HCN line ratio of 9.

Above, we infer a clumpy molecular medium because we chose a $\Delta V$ = 5 \kms,  yielding
small clouds with $r < 1$ pc (cf Aalto \etal 1991). We can not, however, exclude significantly
larger $\Delta V$ which would indicate a continous, non-cloudy structure --- perhaps even a smooth,
rotating disk. A third alternative is a molecular ISM consisting of dense clumps
surrounded by diffuse, non-cloudy molecular gas (e.g. Aalto \etal 1994,1995; Dahmen \etal 1996).

\subsection{The overlap region C and NGC~3690}

Molecular line ratio variations are also seen within the overlap region C, albeit
smaller than those within IC~694.
The weakest \thco\ and strongest HCN 1--0 emission, relative to \twco, is found in C1,
the location of the
brightest H$\alpha$ emission in Arp~299 (Gehrz \etal\ 1983). Continuum emission at
3.4 $\mu$m and 10$\mu$m also peaks close to the C1 clump. Thus, C1 appears to be
currently the most active starforming region within C. We suggest that the observed molecular
line ratio gradients are the result of a temperature and/or density gradient across
the overlap region. The \twco/\thco\ 1--0 line ratio in C1 is considerably lower
than in the nucleus of IC~694, perhaps because C1 is an extranuclear starburst.

Unlike \twco, the \thco\ emission peak is not connected with the nucleus of
NGC~3690 (B1). The radial excitation gradient is similar, therefore, to that of
IC~694, with the highest \twco/\thco\ line ratio in the central region.
The nucleus of NGC~3690, with intense associated H$\alpha$ emission, is a site
of starburst activity (Gehrz \etal 1983).

\subsection{Molecular abundances}

It has been suggested that high \twco/\thco\ intensity ratios in mergers are caused
by unusually high isotopic abundance ratios in molecular clouds with optically
thick \twco\ 1--0 lines. An influx of very low metallicity gas from the
outer disk of the galaxies is the proposed cause of such an extreme abundance ratio
(e.g. Casoli \etal 1992; Henkel \& Mauersberger 1993). Since the \twco/\thco\ intensity
ratio is normal in the outskirts of IC~694 and NGC~3690
this scenario is unlikely to be the explanation for Arp~299. Instead, the measured line ratio
variations most likely indicate differences in the
line excitation and gas properties in different parts of the system.

The observed \twco/\thco\ 1--0 line ratio is, however, a lower limit to the \twco/\thco\
abundance ratio in the emitting region and for $\tau_{10}$(\twco) $\lapprox 1$
this implies an abundance ratio not much greater than 60 in the center of IC~694.
This value is typical for GMCs in the Galactic disk, but higher than in the inner region of our
Galaxy, where the ratio is $\approx 25$ (Langer \& Penzias 1990). Perhaps
the ISM in the nucleus of IC~694 recently arrived from the disk of the galaxy.
In this case, the difference in line ratio between A2 and A1 is solely caused by
a dramatic change in mean optical depth of the \twco\ line.
On the other hand, selective photodissociation of \thco\ by a starburst
and/or an AGN may change the isotopic abundance ratio.
A young nuclear starburst may also produce extra $^{12}$C
and thus temporarily increase the $^{12}$C/$^{13}$C abundance ratio
(e.g., Henkel \& Mauersberger 1993).

\section{Conclusions}

The \twco/\thco\ 1--0 line ratio varies dramatically within Arp~299, from
$60 \pm 15$ at the nucleus of IC~694 to 5-10 in its disk and in
the eastern and north interface regions (C3 and D).
The \thco\ 1--0 brightness, relative to \twco,
is high in quiescent regions of low \twco\ surface brightness, and low
in starburst regions and galactic nuclei. In contrast, HCN 1--0
is bright in the two galaxy centers and in the active extranuclear
star formation region. The \twco/HCN 1--0 is $9 \pm 1$ at the
nucleus of IC~694, $14 \pm 3$ for NGC~3690 and $11 \pm 3$ for the extranuclear
starburst region C1. Unusually high
\twco/\thco\ line ratios ($>$ 20) appear to be associated with extremely
compact molecular
distributions in the nuclei of merging galaxies (cf. Aalto \etal\ 1995).

The large \twco/\thco\ 1--0 intensity ratio at the nucleus of IC~694 can be
attributed to low to moderate optical depth ($\tau \lapprox 1$) in the
\twco\ 1--0 line, possibly combined with abundance effects. A medium consisting
of dense ($n=10^4 - 10^5$ $\cmmd$), warm
($T_{\rm k} > $ 50 K) gas is consistent with the observations.

\acknowledgements
\sskip
We thank Peter Bryant for helpful discussions and suggestions.
The OVRO mm array is supported in part by NSF grant AST 9314079 and
the K.T. and E.L. Norris Foundation.
The National Radio Astronomy Observatory is a facility
of the National Science Foundation operated under
cooperative agreement by Associated Universities, Inc.

\clearpage

\figcaption{(a)$^{12}$CO 1--0 moment map (2800-3400 kms$^{-1}$).
Contour levels are 0.9, 1.8, 3.6, 5.4, 7.2, 9, 10.8,
12.6, 14.4, 16.2, and 18 Jy beam$^{-1}$ \kms. Gray scale levels range
from 5 to 75 Jy beam$^{-1}$ \kms. The peak flux is 97 Jy beam$^{-1}$ \kms
on the nucleus of IC~694. Crosses mark the 6 cm radio continuum positions (Gehrz \etal 1983).
(b) The $^{13}$CO 1--0 moment map. Contour levels are 0.4, 0.8, 1.2, 1.6, 2.0
Jy beam$^{-1}$ \kms. The peak flux is 2.3 Jy beam$^{-1}$ \kms on the disk
of IC~694.(c)HCN 1--0 moment map. Contour levels are 0.6, 1.2, 1.8, 2.4, 3.0, 3.6, 4.2,
4.8, 5.4 Jy beam$^{-1}$ \kms. The peak flux is 7.1 Jy beam$^{-1}$ \kms
at the nucleus of IC~694. \label{fig1}}

\figcaption{The $^{12}$CO 1--0 velocity field. The grayscale range
from 2800 (light) to 3300 (dark) kms$^{-1}$, the contours from 2800 to
3200 kms$^{-1}$. \label{fig2}}

\def\fs{\hbox{$.\!\!^{\rm s}$}}
\def\farcs{\hbox{$.\!\!^{\prime\prime}$}}

\begin{deluxetable}{lccccrr}
\tablecaption{Arp 299:\ \ Source properties}
\tablecolumns{7}
\tablewidth{7.2in}
\tablehead{
\colhead{}& \colhead{}& \multicolumn{2}{c}{$^{12}$CO 1--0 Intensity}& \colhead{}&
    \colhead{}& \colhead{}\nl
\cline{3-4} \\
\colhead{Region}&
        \colhead{Size\tablenotemark{a}}&
        \colhead{Peak}&
        \colhead{Integrated}&
        \colhead{$\Delta V_{\rm FWHM}$}&
    \colhead{${\hbox {\strut $^{12}$CO 1--0}} \over {\hbox {\strut $^{13}$CO 1--0}}$}&
    \colhead{${\hbox {\strut $^{12}$CO 1--0}} \over {\hbox {\strut HCN 1--0}}$}\nl
\colhead{}&
        \colhead{[arcsec]}&
        \colhead{\strut$\left[\hbox{\strut Jy\,km\,s$^{-1}$}\atop\hbox{\strut
            beam$^{-1}$}\right]$}&
        \colhead{[Jy\,km\,s$^{-1}$]}&
        \colhead{[km\,s$^{-1}$]}& \multicolumn{2}{c}{(note c)}}
\startdata
IC\,694: nucleus (A1) &
$\lapprox 1.4$ &   97 &  122 & 350& $60\pm 15$& $9\pm 1$\nl
IC\,694: disk (A2) &
$17 \times \phn 6$\tablenotemark{b} &  12 & 120 & \phn 65& $10 \pm 3$& $>25\ (3\sigma)$\nl
NGC\,3690: nucleus (B1) &
$\lapprox 1.5$ &  29 & \phn 36 & 260& \nodata& \nodata\nl
NGC\,3690: total &
\nodata &  \nodata &  \phn 53 & \nodata& $13\pm 2$& $14\pm 3$\nl
Overlap east (C3)&
$3.4 \times 2.2$ & \phn 9 &  \phn 15 & \phn 60& $10 \pm 3$& $>20\ (3\sigma)$\nl
Overlap west (C2)&
$4.3 \times 2.1$ &  20 & \phn  40 & \phn 80& \nodata& \nodata\nl
Overlap west (C1)&
$3.9 \times 2.0$ &  21 & \phn  47 & \phn 60 & $20 \pm 4$& $11\pm 3$\nl
Region D&
$\dots$ &  $\dots$ & \phn\phn  4 & \phn 80& $5 \pm 3$ & $>10$\nl
Region F&
$\dots$ &  $\dots$ & \phn  11 & 40-80& \nodata& \nodata
\enddata
\tablenotetext{a}{FWHM from two dimensional Gaussian fits using the AIPS task
IMFIT.}
\tablenotetext{b}{An estimate of the size of the SE disk of IC 694 ---
no Gaussian fit possible.}
\tablenotetext{c}{All ratios are in terms of integrated brightness
temperatures.
Uncertainties include thermal noise only. The \twco\ map was smoothed to the
resolution of the \thco\ or HCN map before line ratios were
constructed.}

\end{deluxetable}

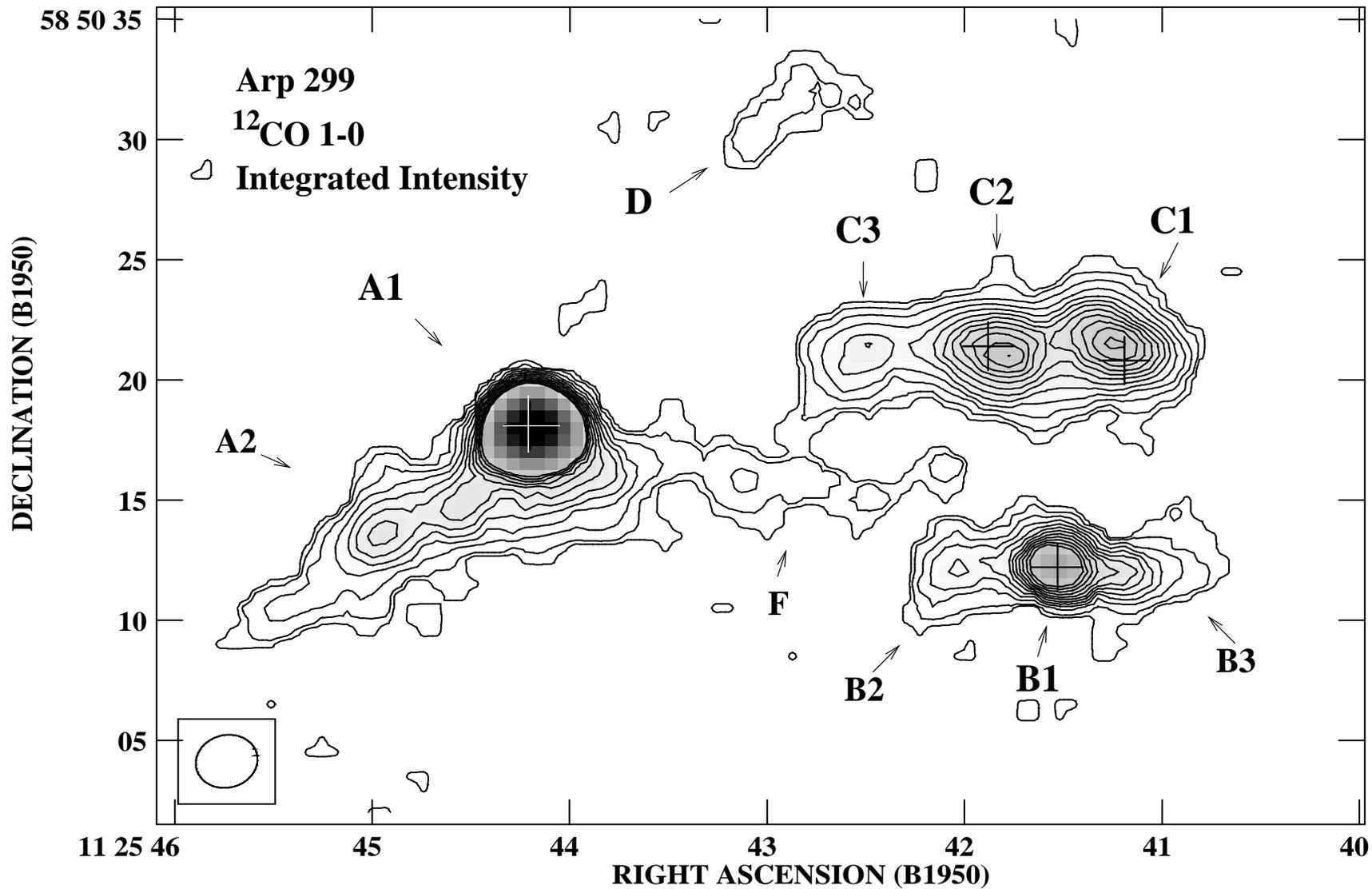
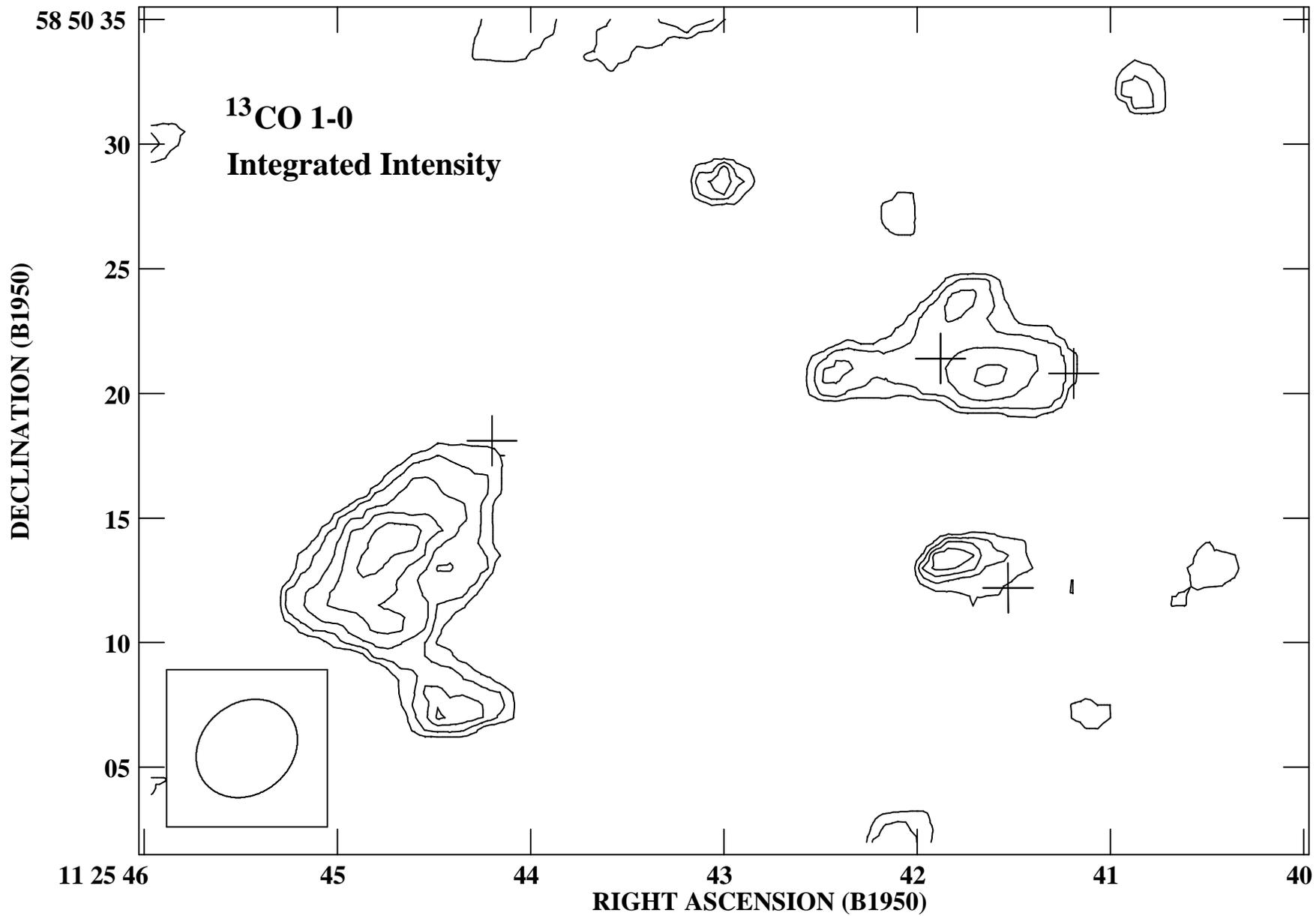
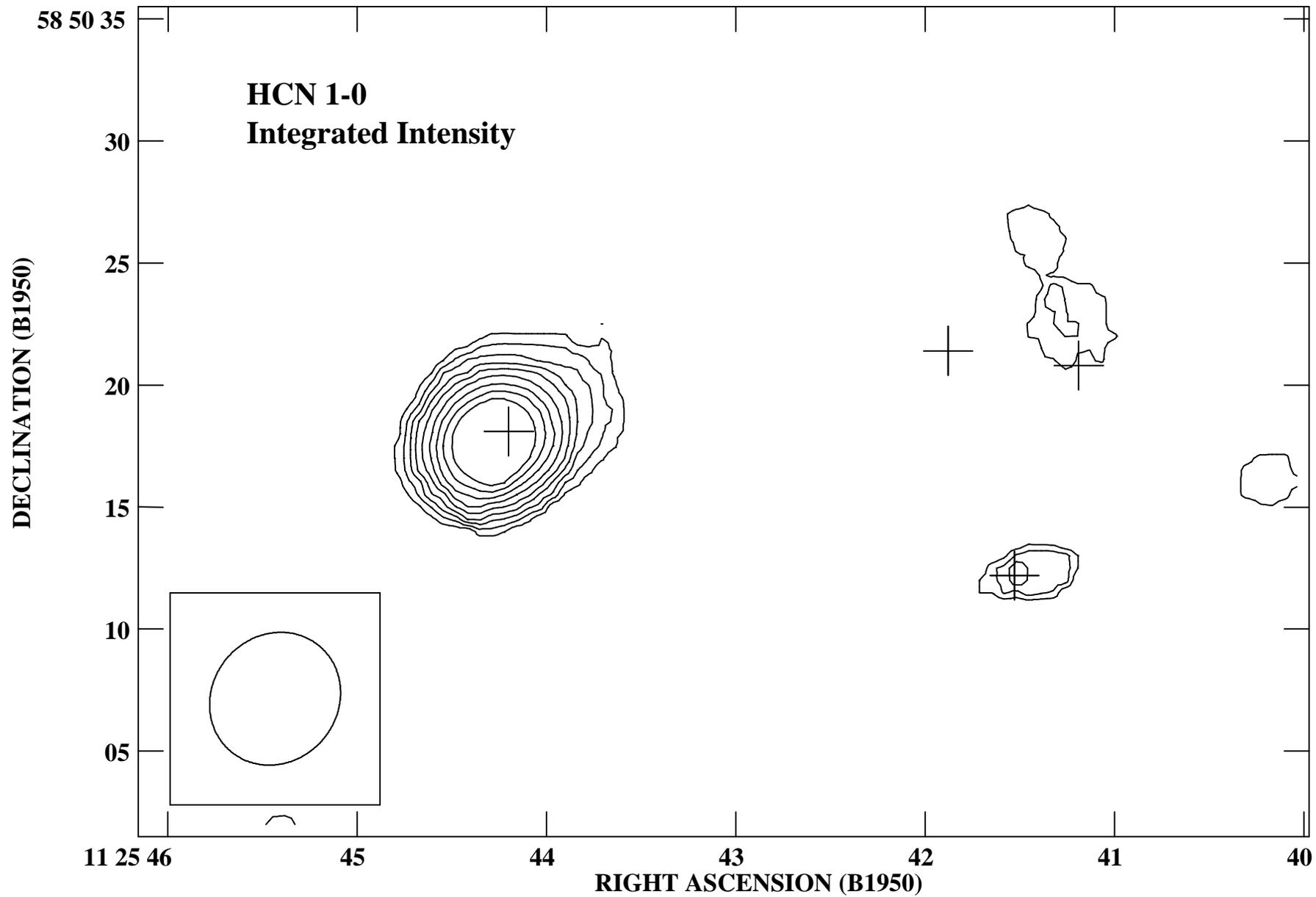
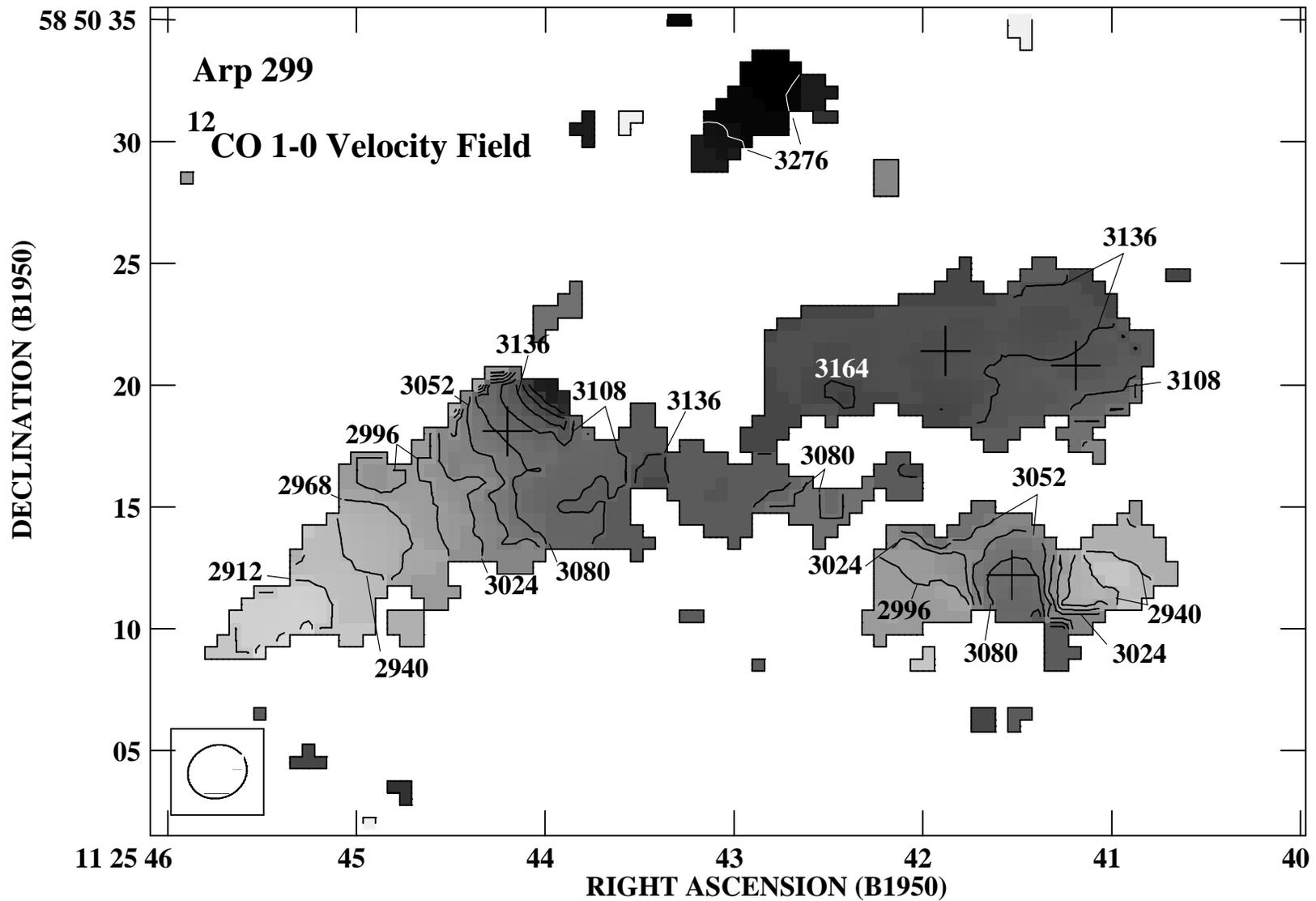
\end{document}